\documentstyle[12pt,epsf]{mrsproc}
\begin{document}
\newcommand{\vareps}{\varepsilon}  
\newcommand{\De}{$\Delta$}
\newcommand{\mc}{\multicolumn}
\begin{center}
{\bf
FIRST-PRINCIPLES CALCULATIONS OF
POSITRON ANNIHILATION IN SOLIDS
}
\end{center}

\vspace{-0.1cm}
B. Barbiellini$^*$, 
M. Hakala$^{**}$, R. M. Nieminen$^{**}$,
M. J. Puska$^{**}$\\

\noindent
$^*$Physics Department,
    Northeastern University, 
    Boston, Massachusetts 02115 \\
$^{**}$Laboratory of Physics,
 Helsinki University of Technology,
 02150 Espoo, Finland \\
 
\vspace{-0.6cm}
\begin{abstract}
\vspace{-0.4cm}
We present first-principles approaches based on density functional
theory for calculating positron states and annihilation characteristics 
in condensed matter. The treatment of the electron-positron correlation 
effects (the enhancement of the electron density at the positron with
respect to mean-field density) is shown to play a crucial role when 
calculating the annihilation rates. A generalized gradient approximation 
(GGA) takes the strong inhomogeneities of the electron density in the ion 
core region into account and reproduces well the experimental total 
annihilation rates (inverses of the positron lifetimes) by suppressing 
the rates given by a local density approximation (LDA). The GGA combined
with an electron-state-dependent enhancement scheme gives a good
description for the momentum distributions of the annihilating 
positron-electron pairs reproducing accurately the trends observed 
in the angular correlation (ACAR) or Doppler broadening measurements of the
annihilation radiation. The combination of the present positron lifetime 
and momentum density calculations with the corresponding measurements
yields a unique tool for defect 
identification. Especially, the investigation of various vacancy-type 
defects in semiconductors able to trap positrons will be an important
field for these methods.
We will show that the identification of vacancy-impurity complexes 
in highly n-Type Si and the study of the SiO$_2$/Si interface
are particularly interesting applications.
\end{abstract}

\vspace{-0.5cm}\section{INTRODUCTION}

\vspace{-0.1cm}
Lattices of crystalline materials are not perfect on the atomic 
scale since point defects always exist. Defects can
dramatically change physical properties of materials, 
even if their concentration is very low.
For instance, in semiconductors defects may increase absorption 
or emission of light in radiative recombination processes.
Experimental methods based on positron annihilation
\cite{krause,review_tdft} can identify point defects in semiconductors 
in concentrations as low as 0.1 ppm. Because the energy-loss 
cross sections for positrons are high, they
slow down to thermal energies before annihilating,
even though they are formed with high energies in beta decay.
The positron lifetimes in solids are typically about 100 - 300 ps. 
If the positron is localized in a vacancy-type defect,
its lifetime increases from that of an extended positron state.
Also the total momentum of the annihilation photon gives useful 
(chemical) information about the environment where the annihilation occurs.
Thus these changes in the positron annihilation characteristics make 
the positron annihilation spectroscopies powerful to study defects in
semiconductors from the viewpoint of material science as well as from that of 
industrial applications. The goal of our work is to develop computational 
approaches for predicting electron and positron states and annihilation
characteristics in materials in order to support the experimental 
research of defects by different positron annihilation methods.
We have have developed models within Density Functional Theory (DFT).

\vspace{-0.5cm}\section{METHOD}

\vspace{-0.1cm}
DFT solves for the electronic structure of a condensed matter
system in its ground state so that the electron density
$\rho^-$ is the basic quantity \cite{review_dft}.
The DFT can be generalized to positron-electron systems by including
the positron density $\rho^+$ as well; it is then called as the 
2-component DFT \cite{review_tdft,tdft}. As a consequence of the 
Hohenberg-Kohn theorem \cite{review_dft} the ground-state value of any 
operator $\hat{o}$ is a functional of the electron and positron 
densities denoted by $O[\rho^-,\rho^+]$.
It can be shown \cite{bauer} that 
\begin{equation}
O[\rho^-,\rho^+]=O_0[\rho^-,\rho^+] + 
\frac{d}{d\lambda} E_{xc}[\rho^-,\rho^+](\lambda)~,
\end{equation}
where $O_0[\rho^-,\rho^+]$ is the expectation value of  $\hat{o}$
for a system of noninteracting fermions moving
in the effective field provided by the Kohn-Sham formalism \cite{review_dft},
$E_{xc}$ is the exchange-correlation energy functional and
$\lambda$ is a scalar coupling parameter for the operator $\hat{o}$.
This general expression for $O[\rho^-,\rho^+]$ generalizes
the Lam-Platzman theorem \cite{bauer} and provides a formal scheme
to extract positron annihilation characteristics from the
2-component DFT.
The Local Density Approximation (LDA) was the first implementation
of the DFT \cite{review_tdft,review_dft} and it provides 
an explicit formula for $E_{xc}$.
The more recent Generalized Gradient Correction (GGA) 
gives a systematic improvement for first-principles
electronic calculations with respect to the LDA \cite{perdew}. 
The GGA is even more successful for positron-electron 
correlation effects in solids when positron lifetimes, energetics, 
and momentum distributions of the annihilating electron-positron pairs are 
considered [7-10]. 

\vspace{-0.5cm}\section{RESULTS}

\begin{figure}[htb]
\unitlength=1cm
\begin{center}
\begin{picture}(8,8.3)
\put(-2.0,-1.0){\epsfysize=9cm
\epsffile{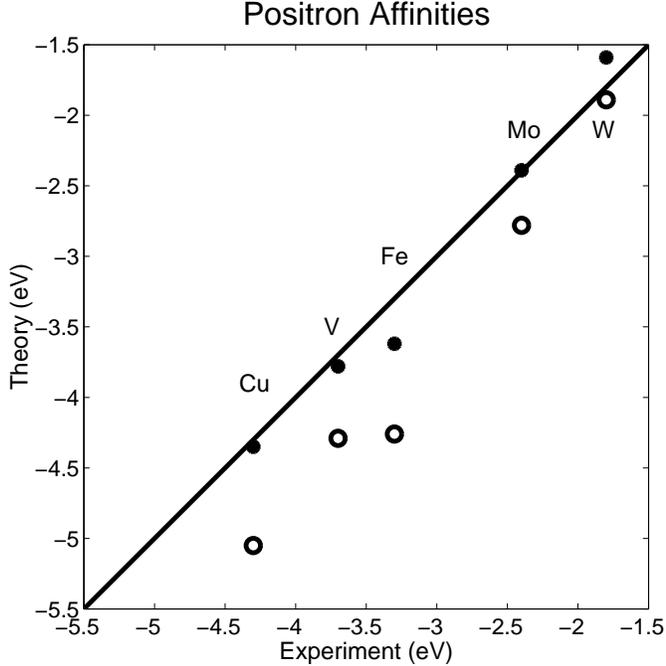}}
\end{picture}
\end{center}
\vskip 0.2cm
\caption{ Positron affinities for several metals. 
The solid and open circles give the GGA and LDA results as a 
function of the experimental ones, respectively. The solid line
corresponds to the perfect agreement between the theoretical and
experimental results from refs. \protect \cite{mills}. }
\label{aff}
\end{figure}

\vspace{-0.1cm}
The positron affinity is an energy quantity defined by $A^+=\mu^- +\mu^+$,
where $\mu^-$ and $\mu^+$ are the electron and positron chemical potentials, 
respectively \cite{boev}. The affinity can be measured by positron 
emission spectroscopy \cite{mills} and the comparison of measured 
and calculated values for different materials is a good test for 
exchange-correlation functionals. The LDA shows a clear tendency 
to overestimate the magnitude of $A^+$ \cite{gga1}.
This overestimation can be traced back to the screening effects. 
In the GGA, the value of $A^+$ is improved with respect to experiment 
by reducing the screening charge. In Fig. \ref{aff} we give the calculated 
positron affinities within LDA and GGA to the corresponding
experimental values for several metals.

In the case of a semiconductor $\mu^-$ is taken from the position
of the top of the valence band. Kuriplach {\em et al.} \cite{kuriplach}
calculated $A^+$ for different polytypes of SiC and 
showed that the GGA 
agrees better with the experimental values than the LDA. However
these experimental data are still plagued with significant error
bars of about $0.5$ eV. Therefore more experimental work
is necessary to judge the accuracy of the present GGA
in semiconductors.
One should also note that the computed $A^+$ values 
may also depend significantly on the wavefunction basis set,
{\em i.e.} whether, for example, the atomic-sphere approximation 
\cite{kuriplach} or the pseudopotential-plane-wave approach \cite{panda} is
used.

%

\begin{figure}[htb]
\unitlength=1cm
\begin{center}
\begin{picture}(8,8.3)
\put(-2.0,-1.0){\epsfysize=9cm
\epsffile{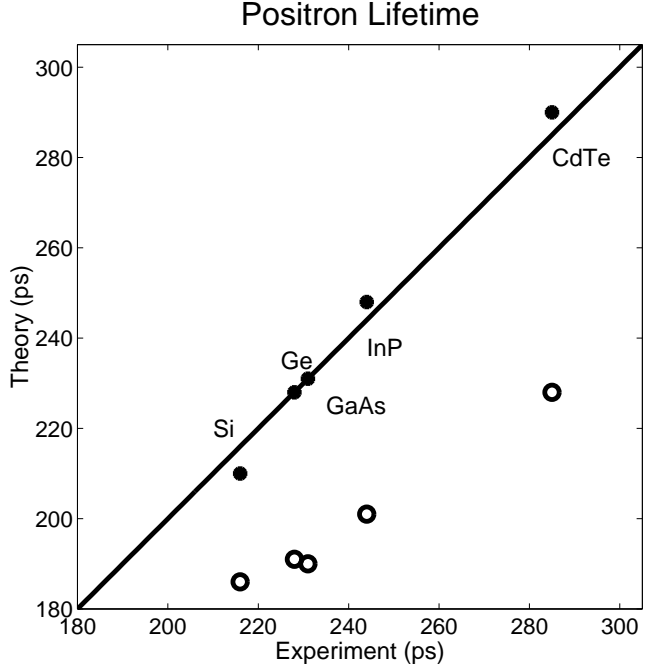}}
\end{picture}
\end{center}
\vskip 0.2cm
\caption{ Positron lifetimes for perfect solid lattices. 
The solid and open circles give the GGA and LDA results as a 
function of the experimental ones, respectively. The solid line
corresponds to the perfect agreement between the theoretical and
experimental results. }
\label{life}
\end{figure}

In a solid, the shape of the screening cloud at the positron 
resembles that of a positronium atom ($Ps$).
The positron-electron contact density, which is remarkably higher than 
the unperturbed electron density, determines the positron lifetime.
The ratio is called the enhancement factor $\gamma$.
In the LDA, the local positron contact density is treated as a function
of the local electron density.
The LDA underestimates systematically the positron lifetime
in real materials while for the GGA, the agreement with
the experiment is excellent as shown in Fig. \ref{life} \cite{gga1,gga2}.
For this comparison we have taken an LDA 
enhancement factor $\gamma$ \cite{gga1} 
which is consistent with the LDA potential given by 
Boronski and Nieminen \cite{tdft}.
Indeed one expects that the strong electric field due to the inhomogeneity
suppresses the electron-positron correlations in the same way
as the Stark effect decreases the contact density for the $Ps$ atom.
In the GGA \cite{gga1,gga2},
the contact density depends also on the density gradient 
and the gradient correction reduces the electron-positron correlation.
The reduction of the enhancement factor due to the GGA is not large in the 
interstitial regions in metals or semiconductors. 
On the contrary, the GGA reduces strongly the enhancement 
factor in the regions of core and semicore electrons.

Very large deviations between theoretical and experimental lifetimes
are observed particularly in II-VI compounds semiconductors, since 
the LDA overestimates the magnitude of the positron annihilation 
rate with the d electrons. Moreover, in the LDA framework, one has 
to introduce for semiconductors and insulators a semiempirical 
correction based on the dielectric constant in order to describe well
the positron lifetime \cite{review_tdft}. This correction is not 
necessary in the GGA. Finally, Ishibashi has shown that the GGA 
reproduces the experimental values much better than the
LDA even for the low-electron-density systems such as the molecular
crystals of C$_{60}$, TTF-TCNQ and (BEDT-TTF)$_2$Cu(NCS)$_2$
\cite{ishibashi}. Therefore the GGA can be safely applied when
identifying vacancy-type defects in semiconductors.


The Doppler broadening technique \cite{alatalo2}
and the two-dimensional angular correlation of
the annihilation radiation (2D-ACAR) \cite{ambigapathy}
are powerful momentum-density spectroscopies for identifying
defects in semiconductors. In the scheme described in Ref. \cite{gga3}
the theoretical momentum distribution is obtained from the shapes which
are determined within the independent particle model for each occupied 
electron 
state and weighted by its annihilation rate calculated within the GGA.
The scheme has been sucessful, for instance, in describing 
the 2D-ACAR spectra of defect-free Cu and GaAs 
\cite{gga3} and the Doppler spectra of compound semiconductors
being defect-free or containing vacancy defects.

The more current theoretical research concerns on the characterization
of defects in Si \cite{si_mikko,si_kimmo}. Momentum densities of 
electron-positron pairs annihilating at vacancy clusters in silicon 
have been calculated \cite{si_mikko}. The theoretical 2D-ACAR spectra
are found to be isotropic if the positron is trapped by a small
vacancy cluster. Calculations indicate that the Doppler-broadened 
lineshape narrows as the size of the vacancy cluster
increases in agreement with experimental data. Moreover, one can 
notice that vacancies and vacancy clusters decorated with impurities 
can produce significantly different lineshapes and therefore the
method can be used in ``chemical analysis'' of defects.
Recently our calculation scheme has allowed us to identify
structures of vacancy-impurity complexes in highly As-doped Si 
samples \cite{si_kimmo}. The comparison of experimental and theoretical
spectra is shown in Fig. [2]. The annihilation with As 3d-electrons raises 
the intensity at high momenta rather linearly as a function of the number 
of As atoms neighboring the vacancy. One can conclude that actually
complexes of a vacancy and three As atoms are formed in as-grown 
highly As-doped Si. Since these complexes are electronically inactive 
they account for the observed saturation of the free electron density.
%

\begin{figure}[th]
\unitlength=1cm
\begin{center}
\begin{picture}(8,8.3)
\put(-2.0,-1.0){\epsfysize=9cm
\epsffile{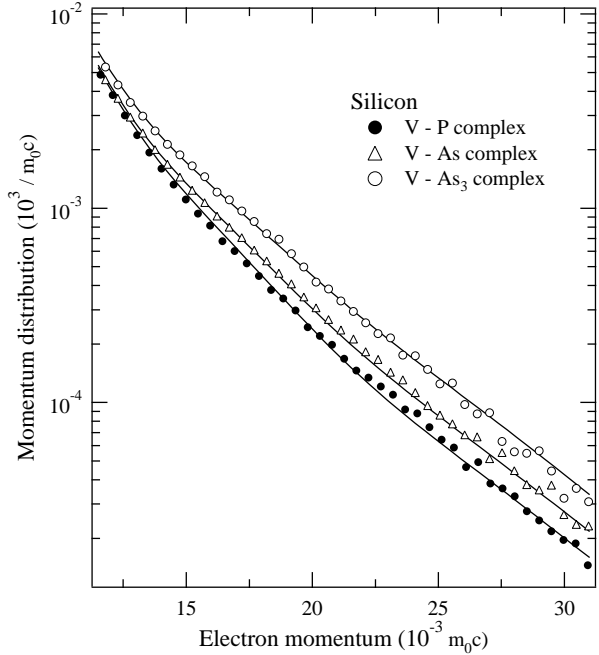}}
\end{picture}
\end{center}
\caption{High-momentum parts of the positron-electron momentum 
distributions  \protect \cite{si_kimmo}. Measured results
for positrons trapped in electron-irradiated P-doped (filled
circles), electron-irradiated As-doped (open circles), and as-grown 
As-doped Si (triangles) samples as well as calculated distributions
(solid lines) for vacancy-P, vacancy-As, and vacancy-As$_3$ complexes 
are shown.
}
\label{Sidop}
\end{figure}

With the same method, one can fingerprint defects on the 
SiO$_2$/Si interface. This interface is very interesting for
the fundamental research since it can realize a two-dimensional 
electron-gas system when gated.
For the past two decades, these systems were believed to be insulating
at low temperatures. However Kravchenko  {\em et al.} \cite{krav} 
have observed an unexpected transition to a conducting phase at
very low electron densities. The nature of this phase is not yet understood.
An interesting hypothesis by Altshuler and Maslov \cite{altshuler}
includes temperature- and field-dependent filling and
emptying of charged hole traps unavoidably introduced during
the device fabrication at the interface.
We are planning to check this hypothesis by comparing the positron
annihilation spectroscopy with our theoretical predictions.
Preliminary results by Kauppinen {\em et al.} \cite{sio}
indeed suggest that high concentrations of
open-volume defects, probably divacancies, exist near the interface.

\vspace{-0.5cm}\section{CONCLUSION}

\vspace{-0.1cm}
We have shown that the scheme based on measuring and calculating  
positron lifetime and momentum distributions is a reliable tool to analyze
materials properties. The study of defect-free bulk samples gives credence 
to use the GGA for the electron-positron correlation effects. The 
localized positron states at defects have also been found well described 
by the GGA. By performing model calculations one can also 
propose new experiments to identify point defects in semiconductors.

\vspace{-0.5cm}\section{ACKNOWLEDGMENTS}

\vspace{-0.1cm}
This work is supported by the US Department of Energy under
Contract No. W-31-109-ENG-38, and benefited from the 
allocation of supercomputer time at the Northeastern 
University Advanced Scientific Computation Center (NU-ASCC).

\newpage

\end{document}